%% file: Odyssey2024_LatexTemplate.tex
\documentclass[a4paper]{article}
\usepackage{Odyssey2024}
\usepackage{epsfig,amssymb,amsmath}
\ninept

\usepackage{graphicx}
\usepackage{amsfonts}
\usepackage{multirow}
\usepackage{booktabs}
\usepackage{array}
\usepackage{enumitem}
\usepackage{adjustbox}
\usepackage{amsmath}

\usepackage{enumitem}
\usepackage{tcolorbox}
\usepackage{enumitem}
\setlist{leftmargin=1mm}
\usepackage{pifont}
\usepackage{setspace}

\usepackage[draft,textsize=footnotesize,textwidth=15mm]{todonotes}

\newcolumntype{P}[1]{>{\centering\arraybackslash}p{#1}}

\title{IMPROVING SPEAKER VERIFICATION ROBUSTNESS WITH SYNTHETIC EMOTIONAL UTTERANCES}

\name{Nikhil Kumar Koditala$^{\star}$  Chelsea Jui-Ting Ju$^{\dagger}$  Ruirui Li$^{\ddagger}$ Minho Jin$^{\mathsection}$ Aman Chadha$^{\bullet}$   Andreas Stolcke$^{\dagger}$}
\address{$^{\star}$Amazon Business, USA  \:\:\: 
$^{\dagger}$Amazon Alexa AI, USA \:\:\:
$^{\bullet}$Amazon GenAI, USA \\
$^{\ddagger}$Amazon Search Experience Science, USA \:\:\: 
$^{\mathsection}$Amazon Web Services, USA 
}

\begin{document}
\maketitle

\begin{abstract}
A speaker verification (SV) system offers an authentication service designed to confirm whether a given speech sample originates from a specific speaker. This technology has paved the way for various personalized applications that cater to individual preferences. A noteworthy challenge faced by SV systems is their ability to perform consistently across a range of emotional spectra. Most existing models exhibit high error rates when dealing with emotional utterances compared to neutral ones. Consequently, this phenomenon often leads to errors in downstream applications, such as authentication systems or emotion recognition applications. 

This issue primarily stems from the limited availability of labeled emotional speech data, impeding the development of robust speaker representations that encompass diverse emotional states.
To address this concern, we propose a novel approach employing the CycleGAN framework to serve as a data augmentation method. This technique synthesizes emotional speech segments for each specific speaker while preserving their unique vocal identity. Our experimental findings underscore the effectiveness of incorporating synthetic emotional data into the training process. The models trained using this augmented dataset consistently outperform the baseline models on the task of verifying speakers in emotional speech scenarios, reducing EER by as much as 3.64\% relative.

\end{abstract}

\section{Introduction}

Speaker verification (SV) systems are designed to discern whether a given audio stream belongs to the speaker who has been previously enrolled, based on various acoustic characteristics of the speaker such as pitch, tone and intonation~\cite{hamid}. SV system typically consists of two phases -- an enrollment phase and a verification phase. During the enrollment phase, the system extracts relevant features from enrolled speaker's speech sample, whereas the verification phase employs statistical or machine learning models to compute distance metric between the current and previously enrolled speech samples~\cite{Bimbot}. The methodologies used for SV modeling have evolved over time. Earlier, techniques such as gaussain mixture models~\cite{940859} and support vector machines~\cite{4291592} are prominent, which are later replaced by neural network based models~\cite{Jung2019RawNetAE, 8462628}. More details on our LSTM-based SV model are presented in Section \ref{svmodel}.

SV models can further be classified into text-dependent and text-independent systems~\cite{Naika}. Text-dependent models aim to verify the speaker based on predefined text, whereas text-independent models are independent of prior textual knowledge. SV systems acts as a key component in various applications such as access control systems, Information structuring, telephonic banking and other voice based authentication systems ~\cite{Bimbot} 
SV models can be used in conjunction with emotion prediction models to recognize particular speech segments and offer tone-related features to individual speakers. These applications encompass scenarios like recognizing individuals from emotional speech during criminal investigations~\cite{nilu}, providing personalized support to individuals experiencing panic or frustration at helpline center~\cite{ nassif2022emotional}, and in wearable emotion recognition devices to monitor psychological health ~\cite{Shu2020WearableER}. 


\begin{table*}[h]
\caption{Data distribution for training and evaluation of SV models. Speech is collected naturally from day-to-day activities, and can thus be a proxy for the general emotion distribution.}
\centering
\vspace{1mm}
\small
\begin{tabular}{cccccccc}
\toprule
\textbf{Dataset} & \textbf{No. of Speakers}& \textbf{No. of Utterances} & \textbf{Neutral} & \textbf{Calm} & \textbf{Angry} & \textbf{Happy} & \textbf{Sad} \\
\midrule
\textbf{Training} & 413 & 132128 & 118948 (90\%) & 3487 (2.6\%) & 751 (0.6\%) & 1344 (1\%) & 7598 (5.7\%) \\
\textbf{Evaluation} & 204 & 55909 & 50860 (91\%) & 1194 (2.1\%) & 494 (0.9\%) & 591 (1.1\%) & 2270 (4.1\%) \\
\bottomrule
\end{tabular}
\label{table:data}
\end{table*}

The advent of SV systems has been a cornerstone in the field of biometric security, offering a seamless method for authenticating individuals in a myriad of settings, from personal devices to secure access control systems. However, the nuanced nature of human speech, particularly when modulated by emotions, presents a formidable challenge that current SV systems are ill-equipped to handle effectively. The variability introduced by emotional states in speech patterns can dramatically affect the accuracy of these systems, leading to potential security vulnerabilities and decreased user satisfaction.

In real-world scenarios, speech is rarely devoid of emotional content. Individuals may express joy, frustration, or stress during interactions with voice-activated systems, leading to discrepancies between the enrolled neutral voice samples and the emotionally charged utterances during verification attempts. This discrepancy is not merely a technical challenge but also a barrier to accessibility and user trust in voice-based authentication technologies. The ability of an SV system to adapt to the emotional variability of speech is not just a feature but a necessity for ensuring the system's reliability and inclusivity.

Moreover, the surge in remote work and digital communications has underscored the importance of robust and adaptable security measures. As voice-based platforms become increasingly integrated into our daily lives, the demand for SV systems that can navigate the complex landscape of human emotions has never been higher. The imperative to address this gap is further amplified by the ethical considerations of fairness and equity in technology deployment. Ensuring that SV systems are insensitive to emotional expressions is critical in preventing unintentional biases and promoting equal access across diverse user groups.

Despite the clear need, the development of SV systems that can effectively process emotional speech is hampered by the scarcity of labeled emotional speech data. This limitation not only restricts the training of more adaptable models but also stifles innovation in creating more human-centric authentication solutions. The pursuit of overcoming these challenges through novel data augmentation techniques, such as the proposed CycleGAN framework, represents a pivotal step towards realizing SV systems that are both secure and empathetic to the human condition.

The integration of emotional intelligence into speaker verification systems is not just an enhancement; it is a critical evolution required to bridge the gap between human speech's dynamic nature and the static models that currently define SV technologies. By addressing the emotional variability in speech, we can unlock new dimensions of security, accessibility, and user experience that align with the nuanced complexities of human communication.

Speakers may intentionally or unintentionally modulate their emotional state during verification, influenced by factors such as stress, urgency, or other physiological factors which may not be present during enrollment~\cite{Purves}. Notably, SV systems often suffer performance degradation when non-neutral tone is present in the speech~\cite{raghavendra}.   
Verification of speakers from emotional utterances is crucial for ensuring fairness and equity in SV systems. Failure to account for emotional variability can lead to inaccurate verification, potentially discriminating against individuals with diverse emotional expressions ~\cite{chien}. Gender related differences also affect the emotional state of speech, since it has been shown ~\cite{swerts} that women experience and perceive emotions more extremely than men. SV on emotional speech can reduce these biases, and ensure equitable access to SV technologies for all users, aligning with the principles of fairness and ethical deployment in biometric systems~\cite{rathgeb}.

Emotional primitives such as activation (indicating high or low energy) and valence (signifying positive or negative feeling) are frequently used to represent different emotions~\cite{Bestelmeyer}. Activation aids in distinguishing between emotional states like angry or happiness (high energy) and sadness or calmness (low energy); valence helps differentiate between angry (negative feeling) and happiness (positive feeling)~\cite{carlos}. 

When a person experiences anger, the speech often deviates from his/her neutral tone,
resulting in a significant \textit{tone mismatch}. This presents a major challenge for the SV systems when comparing such emotionally-inflected speech with the enrollment data. Addressing emotions in speaker recognition is a crucial aspect that has been relatively unexplored. This approach has broader implications for improving the performance of speaker recognition models in various practical applications, especially in real-world scenarios where emotions play a pivotal role in human communication. Moreover, building a robust speaker recognition system is hindered by scarcity of emotional utterances data. Since we speak in a neutral tone most of the time, only a meagre fraction, $\sim$10\% of the data, captures the emotional tone (cf. Table~\ref{table:data}).
Consequently, the error rates (false acceptance and false rejection) of our SV system, which is a multi-layer LSTM network trained using generalized end-to-end (GE2E) loss \cite{liwan} is 1.3\% higher for emotional utterances compared to neutral ones (Table~\ref{tab:synthetic_all}). 

To tackle the data sparsity issue, we propose to use emotional voice conversion techniques to supplement the training data. This involves generating emotional utterances from neutral ones by preserving the speaker's identity, thereby enhancing the effectiveness of training for SV with emotional speech.

Emotional voice conversion transforms audio samples of a specific emotional tone into audio samples of a different emotional tone without changing any linguistic content. 
Existing converters can be categorized into statistical-based or neural network-based approaches. Statistical-based approach includes Gaussian mixture models~\cite{tomoki}, non-negative matrix factorization~\cite{daniel}, and partial least square regression~\cite{zhizheng}. These methods operate on low dimensional spectral features, and their performance is extremely sensitive to the quality of the input speech. Neural network-based approaches encompass restricted Boltzmann machines ~\cite{ling}, feed-forward neural networks~\cite{srinivas}, and artificial neural networks~\cite{nakashika}. These implementations rely on the availability of parallel training data, which means having a dataset containing identical linguistic content spoken by the same speakers, but with various emotional expressions. Collecting such dataset is impractical, as it is not natural for a speaker to repeat the same sentence with 4-5 different emotions. 

In recent developments, several approaches have emerged aiming to circumvent the need of parallel data in style-transfer tasks. These methods include Deep Bidirectional Long-Short-Term Memory (DBLSTM) with i-vector~\cite{7820901}, variational auto-encoder~\cite{8461384,hsu2016voice} and GANs~\cite{2017arXiv170400849H,9003939,8553236}. 
Recently, Zhou \textit{et al.}~\cite{zhou} introduced a CycleGAN network to learn the emotional voice conversion task based on a style transfer autoencoder. Furthermore, the authors explore the application of continuous wavelet transform (CWT) to enhance the efficacy of F0 conversion. This architecture eliminates the need for parallel training data or any external modules, and has emerged as a viable option to synthesize emotional utterances that fit our need.


We have trained two CycleGAN networks that transform neutral-tone utterances into angry-tone and happy-tone utterances while preserving their distinctive speaker characteristics. The synthetic emotional utterances are used to augment the training dataset used to learn the SV task. Results show that models trained with a combination of authentic and synthetic emotional data have demonstrated a relative reduction of 1.08\% to 3.64\% in equal error rate (EER) for emotional utterances.

To summarize, the key contributions of our research can be outlined as follows:


\vspace{-2.5mm}
\begin{tcolorbox}
[colback=black!5!white,colframe=black!75!black,title=\textsc{\normalsize Our Contributions}]
\begin{itemize}
[leftmargin=1mm]
\setlength\itemsep{0em}
\begin{spacing}{0.85}
\vspace{1mm}
    \item[\ding{224}] {\footnotesize 
    {\fontfamily{phv}\fontsize{8}{8}\selectfont To the best of our knowledge, this is the first work that has applied CycleGAN as a data augmentation technique for the SV task. 
    } 
    }
\vspace{2mm} 
    \item[\ding{224}] {\footnotesize 
    {\fontfamily{phv}\fontsize{8}{8}\selectfont The proposed method leads to improved SV performance on emotional speech owing to a reduced performance gap between neutral and emotional utterances. 
    }
    }    
\vspace{-1mm} 
    \item[\ding{224}] {\footnotesize 
    {\fontfamily{phv}\fontsize{8}{8}\selectfont Experimental results have shown that the trained CycleGAN networks are able to effectively preserve unique speaker characteristics during emotion conversion.
    }
    }   

\vspace{-3mm}    
\end{spacing}    
\end{itemize}
\end{tcolorbox}



\section{Methods}

\subsection{Speaker Verification Model}\label{svmodel}

The SV model is a multi-layer LSTM network which uses a 40-dimensional Mel-spectogram as input \cite{fusion} and outputs an $n$-dimensional \textit{d-vector} or \textit{deep vector}, which is the average of activations derived from the last hidden layer of the LSTM~\cite{dvector}. d-vector
helps in speaker verification by encapsulating the neural embeddings of a speaker's voice characteristics. This network is trained using the generalized end-to-end (GE2E) loss proposed by Li \textit{et al.}~\cite{liwan}. 

In the inference phase, the system takes a pair of d-vectors  one representing the speaker profile derived from enrollment data, and the other containing the speaker information of an utterance of interest. We use cosine similarity between these two d-vectors to verify if the pair corresponds to the same speaker. We acknowledge that the baseline model does not represent the state-of-the-art (SOTA). This deliberate choice was made to ensure that the SV model is functional within production-level constraints imposed to work on relatively smaller resources, be it on-device configurations or limited worker farms supporting cloud back-ends.

\subsection{CycleGAN for Emotion Conversion}
The CycleGAN framework uses WORLD Vocoder~\cite{masanori} to extract speech features from an utterance. The emotion converter contains two components: one uses 24-dimensional Melcepstral coefficients (MFCCs) for spectrum conversion, and the other uses 10-dimensional Fundamental frequency (F0) features to handle prosody conversion are fed into a separate CycleGAN network ~\cite{zhou}. 
Similarly to other GAN architectures, CycleGAN incorporates a generator for converting audio features from one emotional tone to another, and a discriminator for discerning the real and the converted data.
We use three different loss functions to train the CycleGAN framework: (i) adversarial loss, (ii) cycle-consistency loss, and (iii) identity loss.

Adversarial loss, defined in Equation \eqref{adv}, measures the distinctness between the converted data and the original data, specifically,
\begin{align}
\label{adv}
L_{A D V}\left(G_{X \rightarrow Y},\right. & \left.D_Y, X, Y\right)=\mathbb{E}_{y \sim P(y)}\left[D_Y(y)\right]\\
\notag
& +\mathbb{E}_{x \sim P(x)}\left[\log \left(1-D_Y\left(G_{X \rightarrow Y}(x)\right)\right]\right.
,
\end{align}
where $x \in X$ is the source utterance (with a neutral tone) and $y \in Y$ is the target utterance (with an emotional tone). $G_{X \rightarrow Y}$ is the Generator that converts source to target utterances, whereas $D_{Y}$ is the Discriminator for target utterances. 

Adversarial loss only governs whether ${G_{X \rightarrow Y}}$ follows the distribution of the target data but does not preserve the contextual information nor the speaker identity. Hence, cycle-consistency loss is used to guarantee the consistency of contextual information and speaker identity between $x$ and ${G_{X \rightarrow Y}}$. This loss encourages the generator to find an optimal pseudo pair $(x,y)$ through circular conversion. Cycle-consistency loss is defined in Equation \eqref{ccloss} as follows, 
\begin{align}
\label{ccloss}
& L_{C Y}\left(G_{X \rightarrow Y}, G_{Y \rightarrow X}\right) \\
& \qquad \begin{array}{l}
\notag
=\mathbb{E}_{x \sim P(x)}\left[\left\|G_{Y \rightarrow X}\left(G_{X \rightarrow Y}(x)\right)-x\right\|_1\right] \\
\quad+\mathbb{E}_{y \sim P(y)}\left[\left\|G_{X \rightarrow Y}\left(G_{Y \rightarrow X}(y)\right)-y\right\|_1\right]
\end{array} \quad
\end{align}
Lastly, we use identity loss to preserve linguistic information. Identity loss is defined in Equation \eqref{iloss} as follows,
\begin{align}
\label{iloss}
& L_{I D}\left(G_{X \rightarrow Y}, G_{Y \rightarrow X}\right) \\
= & \mathbb{E}_{x \sim P(x)}\left[\left\|G_{Y \rightarrow X}(x)-x\right\|\right]+\mathbb{E}_{y \sim P(y)}\left[\left\|G_{X \rightarrow Y}(y)-y\right\|\right]
\notag
\end{align}

The final loss function is a combination of the above three loss functions:
\begin{align}
\label{comb}
& L(G_{X \rightarrow Y},G_{Y \rightarrow X},D_X,D_Y, X, Y)\\
\notag
& = L_{A D V}(G_{X \rightarrow Y}, D_Y, X, Y) + L_{A D V}(G_{Y \rightarrow X}, D_X, X, Y)\\
& + \lambda_{CY} L_{C Y }\left(G_{X \rightarrow Y}, G_{Y \rightarrow X}\right) + \lambda_{ID} L_{I D}\left(G_{X \rightarrow Y}, G_{Y \rightarrow X}\right)
\notag
\end{align}

To guide the learning process, we set $\lambda_{CY}=10$ for all the iterations, whereas $L_{ID}$ is only used for the first $10^4$ iterations with $\lambda_{ID}=5$.



\section{Experiment Settings}

\subsection{Data}
To train the CycleGAN model, we used non-parallel emotional utterances drawn from three distinct open source datasets: Emotional Speech Dataset \cite{kun}, EmoV \cite{adigwe}, and Ravdess \cite{livingstone}. Emotional Speech dataset consists of 350 utterances spoken by 10 English speakers in different emotional states. EmoV dataset is a collection of utterances from 4 different speakers. Ravdess is an audio-visual dataset containing 1440 emotional utterances spoken by 24 professional actors.

The aforementioned data is abundant in emotional content but lacks a sufficient number of speakers.
For SV models, we acquired recordings from everyday life through internal channels with proper consent. To be compliant with biometric regulations like General Data Protection Regulation (GDPR), California Consumer Privacy Act (CCPA), and Biometric Information Privacy Act (BIPA), we obfuscate the biometric data pertaining to our speakers.
In order to generate the speaker labels, 
 we first randomly sampled anonymized utterances. These selected utterances, along with the enrollment data corresponding to the same device's speakers, were then provided to multiple annotators, with a minimum of 3 annotators independently assigning ground-truth labels. To minimize annotation errors, we adopted the decision agreed upon by the majority of annotators.
We divided the data into training and evaluation subsets, with distinct sets of speakers. Within the training set, we randomly set aside 5\% of the speakers as our validation set for early stopping~\cite{Prechelt}. Table ~\ref{table:data} provides an overview of the distribution of speakers and utterances in the training and evaluation sets. 


\begin{table}[b]
\caption{Average cosine similarity between neutral tone and angry tone utterances of the same speaker.}
\vspace{2mm}
\adjustbox{max width=\columnwidth}{%
\small
\begin{tabular}{p{4cm}P{1.7cm}P{1.7cm}}
\toprule
\multirow{2}{*}{\textbf{Case}}
 & \multicolumn{2}{c}{\textbf{Cosine Similarity}} \\
 
 & \textbf{Speaker 1} & \textbf{Speaker 2} \\
\midrule

 Neutral vs. Authentic Angry & $0.51 \pm 0.10$ & $0.53 \pm 0.12$ \\
 
 Neutral vs. Synthetic Angry & $0.65 \pm 0.06$ & $0.65 \pm 0.09$ \\
\bottomrule
\end{tabular}}
\label{table:cosine}
\end{table}

\input{synthetic_utterance_performance_192spkr.tex}

\input{synthetic_utterance_performance_all.tex}

\subsection{Model Training}
\subsubsection{Training Considerations}\label{tcons}
We trained two separated CycleGAN networks, each dedicated to transforming neutral-tone utterances into either angry or happy tones. 2000 audio utterances per emotion are sampled from three different datasets i.e. Emotional Speech Dataset \cite{kun}, EmoV \cite{adigwe}, and Ravdess \cite{livingstone} are used to train the model. During the training phase of the CycleGAN network, the input consists of source (neutral) and target (emotional) utterances from same speaker, albeit containing different emotional expressions and even distinct linguistic content. The CycleGAN model concurrently learns forward and inverse mapping with the help of adversarial and cycle consistency losses. This approach aims to ascertain the optimal transformation between source (neutral) and target (emotional) utterances. We focus on synthesizing the angry- and happy-tone utterances because these two emotional categories exhibit high energy levels and are situated farther apart from the neutral tone within the Activation-Valence spectrum.

Throughout the adversarial training process, the generator and discriminator are competing with each other to learn the network. We expect the discriminator loss to stabilize, which means that the discriminator is not able to differentiate correctly between the generated samples and the real samples. However, the discriminator task is relatively easy at the beginning of the training since the generator has not learned to generate adequate data. This made the generator fail due to vanishing gradients ~\cite{arjovsky2017towards}. 

To mitigate this issue, we adopt a strategy that allows the generator a head start in learning~\cite{ian}. This method involves providing an initial advantage to the generator by refraining discriminator from training for the first $k$ train steps. With this technique, the generator is afforded a distinct oppurtunity to undergo training and improve its performance without immediate interference from the discriminator. Consequently, this approach facilitates generator's advancement ahead of discriminator's active engagement, thereby mitigating the disparity in learning progress. 

Similarly, we observed an improvement of F0 model performance by using regularization techniques such as early stopping. We noted a $16.5\%$ improvement in consine similarity between source and prediction embeddings of same speaker.

\subsubsection{Training Configuration and Hyperparameters}
We employed the generalized
end-to-end (GE2E)~\cite{liwan} training framework, where we assembled a mini-batch consisting of $N \times M$ utterances. Specifically, these utterances were gathered from $N=32$ speakers, and each speaker contributes $M=5$ utterances. The features extracted from each of the training utterances $x_{ij}$ $(1 \leq j \leq N$ and $1 \leq i \leq M )$ are fed into an LSTM network. 

During the training process, the GE2E loss pushes the centroid of embeddings closer to the true speaker and away from negative nearby speakers. We utilized the Adam optimizer~\cite{adam} for training the model parameters, with a learning rate $\eta = 10^{-6}$, and an exponential decay rate $\beta$ = 0.98 at every 10k iterations.

\section{Results}

\subsection{Emotion Conversion}
\begin{figure}[t]
\centerline{\includegraphics[scale=0.25]{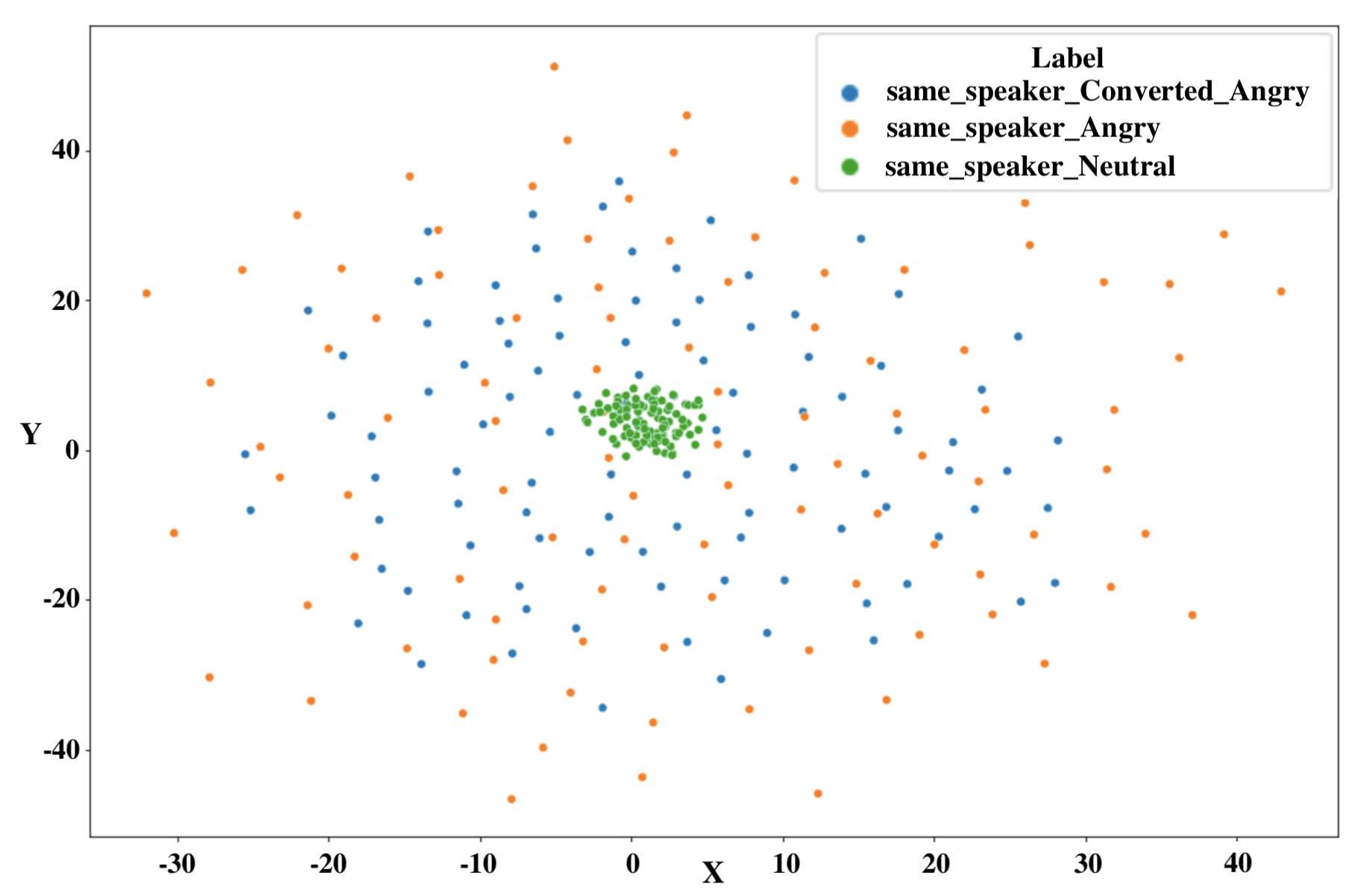}}
\caption{t-SNE plot of utterances from a single speaker in neutral tone (\textcolor{teal}{green}), angry tone (\textcolor{blue}{blue}), and converted angry tone (\textcolor{orange}{orange}).}
\label{t-SNE}
\end{figure}

As discussed in Section \ref{tcons}, we trained two different CycleGAN networks to transform neutral-toned utterances into those expressing anger and happiness. To assess the efficacy of this emotional modulation, we conducted a comprehensive evaluation. Firstly, we utilized the cycleGAN model to synthesize angry utterances from neutral utterances. We then extracted speaker embeddings using the baseline SV model for three distinct sets of utterances (i) neutral tone, (ii) angry tone, and (iii) a converted angry tone via the CycleGAN network.
We then used t-SNE~\cite{tsne_visual} to project these embeddings into a 2-dimensional space for visualization. To enhance the readability of the visualization, we opted to utilize data from a single speaker exclusively. 
Figure~\ref{t-SNE} shows that neutral utterances form a central cluster, while the angry-tone utterances scattered around the space. Notably, the synthetic utterances seamlessly integrate with the authentic angry tone utterances, indicating a change in the emotional state of the utterances.


Since the primary application of these synthetic emotional utterances lies in training SV models, a critical aspect of the converter is preserving the speaker's vocal characteristics. To evaluate the speaker information, 
we measure the cosine similarities based on speaker embeddings between the neutral and angry-tone utterances. Specifically, we computed the average and standard deviation of cosine similarities between neutral and real angry-tone utterances of the same speaker. This serves as a baseline for the expected similarity between neutral- and angry-tone utterances of a single speaker. We then computed similar metrics between utterances in neutral and synthetic angry-tone of the same speaker. 
Table~\ref{table:cosine} shows that synthetic angry utterances exhibit higher cosine similarities with the neutral tone utterances compared to the authentic angry utterances. This suggests that our emotion converter effectively retains unique speaker characteristics.


\subsection{Augmenting Emotional Data to Train the SV Model}
In this section, we explore the impact on the performance of the SV model by augmenting the training dataset with converted emotional data. In our preliminary investigation, we sub-sampled 192 active speakers from our training dataset that have a sufficient number of neutral utterances. These neutral utterances are fed into the WORLD Vocoder~\cite{masanori} to extract F0 and MFCC features. We leveraged the trained CycleGAN models to convert the existing neutral prosody and spectrum features to synthesize angry- and happy-tone utterances. 

The baseline model is trained with 50 neutral-tone utterances collected from each of these 192 speakers. The experimental models are trained with both neutral-tone (real) and emotional (synthesized) utterances in different proportions. We trained four experimental models using synthetic angry utterances: (i) \textit{50 neutral + 10 angry}, (ii) \textit{50 neutral + 20 angry}, (iii) \textit{50 neutral + 50 angry}, and (iv) \textit{60 neutral + 20 angry}, where the notation ``\textit{50 neutral + 10 angry}" signifies that within our training dataset, consisting of 192 speakers, we incorporated 50 neutral utterances along with 10 synthetic angry utterances for each speaker. Similarly, we trained experimental models by including synthetic happy utterances, viz. configuration (i) \textit{50 neutral + 10 happy} and (ii) \textit{50 neutral + 20 happy}, as well as experimental models including both synthetic happy and angry utterances as shown in Table~\ref{tab:synthetic_192}.

These models are evaluated on the same evaluation set described in Table~\ref{table:data}, and we evaluate the overall performance together with the performance breakdown on different emotion categories. Table~\ref{tab:synthetic_192} summarizes the relative reduction on EER (EER) of the SV models trained with the inclusion of synthetic data. We would like to highlight that in accordance with the internal guidelines of our organization, we are restricted from disclosing the absolute EER. Therefore, we have presented the relative EER as an alternative measure.


Overall, models trained with the inclusion of synthetic angry- and happy-tone data have effectively reduced errors for emotional utterances (happy, angry, sad, and calm). However, this improvement is accompanied by a decline in performance for neutral utterances, especially when synthetic angry-tone data is introduced. For instance, the *50 neutral + 20 angry* configuration has led to an improvement in EER of 3.00\% on emotional utterances, but a degradation of -0.65\% on neutral utterances from the evaluation dataset. As we progressively increase the number of synthetic angry-tone utterances, we observe a boost in performance for angry-tone utterances, but a consistent degradation for neutral utterances. Table~\ref{tab:synthetic_192} shows that as we increase the synthetic angry-tone proportion from 10, 20, to 50 per speaker, we observed a decline in neutral utterances from -0.28\%, -0.65\% to -3.77\%. Yet, such degradation is less pronounced when increasing the number of synthetic happy-tone utterances. By adding 10, 20 synthetic happy utterances we noticed an improvement in overall EER by 0.61\% (0.69\% in neutral and 0.86\% in emotional) and 1.44\% (2.02\% in neutral and 4.57\% in emotional) respectively. This suggests that adding synthetic happy-tone data is more effective in learning the SV task compared to angry-tone data.


There are two potential factors contributing to the observed performance gain. One possibility is that the synthetic data provides genuine signals to learn the SV task. The other possibility is the increase in data size. To disentangle these influences, we conducted an experiment with more utterances per speaker (cf. \textit{60 neutral + 20 angry} configuration in Table~\ref{tab:synthetic_192}). 
If data size were the primary factor, we would expect better performance on this setting compared to others trained with fewer utterances. Contrary to this expectation, the overall performance was only superior to that of the \textit{50 neutral + 20 angry} configuration in Table~\ref{tab:synthetic_192}. Similarly, when we introduced a substantial number of synthetic angry-tone utterances (cf. \textit{50 neutral + 50 angry} configuration in Table~\ref{tab:synthetic_192}), the model did not outperform others with fewer utterances. These findings suggest that data size is not the predominant factor, and synthetic data contains valuable signals that assist the learning of the SV task.

Next, we delve into the impact of training the SV models with a combination of authentic and synthetic emotional data on the entire training dataset. We incorporate utterances from all speakers. As presented in Table~\ref{tab:synthetic_all}, the addition of a modest quantity of synthetic data (cf. \textit{5 angry + 5 happy} configuration in Table~\ref{tab:synthetic_all}) leads to a relative error reduction of 0.81\% for neutral utterances, and 1.08\% for emotional utterances. Consistent with our previous findings, as we incrementally augment the number of synthetic emotional utterances, we continue to witness improvements for emotional utterances, albeit with a slight degradation for neutral utterances. This approach of including synthetic emotional utterances enables us to narrow the error gap between neutral and emotional utterances, reducing it from 1.30\% to 0.94\%.

\subsubsection{Resilience against Spoofing Attack}
Integration of synthetic utterances as a data augmentation technique presents risk of spoofing attacks ~\cite{10096733}, which involves malicious use of audio utterances to deceive the SV model. Here, we specifically focused on media speech as a proxy for spoofing, given its prevalence in everyday communication scenarios. The term ``media speech" pertains to audio recordings sourced from various mediums such as television broadcasts or YouTube videos. It is crucial for SV models to differentiate media speech content from explicit user commands. Owing to our utilization of synthetic data for training, investigating whether this practice has had any adverse impact on the performance of SV models when applied to media speech data was necessary.


For this experiment, we considered 3000 media speech utterances and we compared each of them against all the speakers in the evaluation data set. Despite adding synthetic utterances, we observed that the FAR value on media speech is less than the targeted $3\%$ FAR in both Neutral and Emotional spectrum, which proves that adding synthetic data did not negatively affect the performance on media speech. Nonetheless, further research is imperative to fortify SV models against evolving spoofing techniques.

\section{Conclusion}

In this work, we have introduced a pioneering approach utilizing the CycleGAN framework to significantly improve SV systems through the innovative use of data augmentation. This technique uniquely maintains the individual vocal traits of speakers while generating synthetic emotional speech samples, thus enhancing the training process and improving model performance across various emotional scenarios. This integration of synthetic data marks a pivotal improvement in the accuracy and reliability of SV models, addressing the critical need for systems that can adeptly handle the nuances of human speech influenced by emotional states.

The implications of this advancement are profound, offering new possibilities for creating synthetic data that closely mimics a wide range of acoustic conditions, from the specific properties of different recording devices to the nuances between near-field and far-field audio, and even the subtle differences in speaker gender. This opens the door to significantly improved speaker verification systems that can more accurately and effectively authenticate users under diverse conditions, making these systems more versatile and powerful tools for security and identification purposes.

Furthermore, our findings underscore the importance of emotional variability in the context of SV systems, highlighting how addressing this variability directly contributes to the systems' robustness, adaptability, and fairness. By enhancing the ability of SV systems to accurately recognize and verify speakers across a broad spectrum of emotional expressions, we ensure more equitable access to authentication technologies. This focus on emotional intelligence within biometric systems aligns with an increasing demand for security solutions that are both advanced and attuned to the complexities of human behavior, ensuring a more inclusive approach to authentication that respects and accommodates the diversity of user populations.

Looking ahead, the potential applications of our CycleGAN-based data augmentation extend beyond the immediate scope of speaker verification. The same principles could be applied to enhance voice interaction systems in smart environments, support empathetic responses in AI-driven customer service, and even contribute to advancements in forensic analysis by enabling more accurate speaker identification under varied emotional conditions. The versatility of our approach paves the way for more empathetic and responsive AI systems and opens new avenues for research and development aimed at bridging the gap between current static models and the dynamic needs of real-world communication.




\section{Acknowledgments}
We would like to thank Brecht Desplanques for his contribution in reviewing this work and providing valuable feedback. 

\bibliographystyle{IEEEbib}
\bibliography{Odyssey2024_BibEntries}

%

\end{document}

%% file: synthetic_utterance_performance_192spkr.tex
\begin{table*}[h]
\centering
\caption{Relative improvement in EER comparing the SV models trained with neutral utterances only (baseline) and with addition of synthetic angry and happy utterances. Training data contains sub-sampled speakers, with each speaker contributing 50-60 neutral utterances. Positive values signify an improvement in performance compared to the baseline.}
\vspace{0.2cm}
\setlength{\tabcolsep}{4pt}
\small
\begin{tabular}{llccccccc}
\toprule
\multicolumn{2}{c|}{\textbf{Experiment Configuration}}                                                                                                                         & \multicolumn{1}{c|}{\textbf{Overall}} & \textbf{Neutral}     & \multicolumn{1}{c|}{\textbf{Emotional}} & \textbf{Happy}       & \textbf{Angry}       & \textbf{Sad}         & \textbf{Calm}        \\ 
\midrule
\textbf{Baseline}                                                                                       & \multicolumn{1}{l|}{50 neutral}                          & \multicolumn{1}{c|}{-}                & -                    & \multicolumn{1}{c|}{-}                  & -                    & -                    & -                    & -                   \\ \midrule
\multirow{4}{*}{\textbf{\begin{tabular}[c]{@{}l@{}}Synthetic angry \\ utterances\end{tabular}}}         & \multicolumn{1}{l|}{50 neutral + 10 angry}                  & \multicolumn{1}{c|}{0.74\%}           & -0.28\%              & \multicolumn{1}{c|}{0.29\%}             & 4.38\%               & -7.58\%              & -1.46\%              & 3.90\%               \\
                                                                                                        & \multicolumn{1}{l|}{50 neutral + 20 angry}                  & \multicolumn{1}{c|}{-0.14\%}          & -0.65\%              & \multicolumn{1}{c|}{3.00\%}             & 10.18\%              & 2.69\%               & -1.17\%              & 5.54\%               \\
                                                                                                        & \multicolumn{1}{l|}{50 neutral + 50 angry}                  & \multicolumn{1}{c|}{-1.74\%}          & -3.77\%              & \multicolumn{1}{c|}{1.64\%}             & -1.42\%              & 21.61\%              & -5.12\%              & 6.45\%               \\
                                                                                                        & \multicolumn{1}{l|}{60 neutral + 20 angry}                  & \multicolumn{1}{c|}{0.57\%}           & -0.20\%              & \multicolumn{1}{c|}{2.50\%}             & 11.60\%              & -1.63\%              & -2.34\%              & 3.90\%               \\ \midrule
\multirow{2}{*}{\textbf{\begin{tabular}[c]{@{}l@{}}Synthetic happy \\ utterances\end{tabular}}}         & \multicolumn{1}{l|}{50 neutral + 10 happy}                  & \multicolumn{1}{c|}{0.61\%}           & 0.69\%               & \multicolumn{1}{c|}{0.86\%}             & -1.42\%              & 7.02\%               & -1.46\%              & 0.00\%               \\
                                                                                                        & \multicolumn{1}{l|}{\textbf{50 neutral + 20 happy}}         & \multicolumn{1}{c|}{\textbf{1.44\%}}  & \textbf{2.02\%}      & \multicolumn{1}{c|}{\textbf{4.57\%}}    & \textbf{0.00\%}      & \textbf{7.02\%}      & -0.59\%              & \textbf{3.90\%}      \\ \midrule
\multirow{3}{*}{\textbf{\begin{tabular}[c]{@{}l@{}}Synthetic angry + \\ happy utterances\end{tabular}}} & \multicolumn{1}{l|}{50 neutral + 5 angry + 5 happy}            & \multicolumn{1}{c|}{0.72\%}           & -0.71\%              & \multicolumn{1}{c|}{2.57\%}             & -1.42\%              & 4.32\%               & -0.29\%              & -1.36\%              \\
                                                                                                        & \multicolumn{1}{l|}{\textbf{50 neutral + 10 angry + 10 happy}} & \multicolumn{1}{c|}{\textbf{1.61\%}}  & \textbf{1.30\%}      & \multicolumn{1}{c|}{\textbf{2.35\%}}    & \textbf{4.38\%}      & \textbf{1.07\%}      & -0.59\%              & \textbf{3.90\%}      \\
                                                                                                        & \multicolumn{1}{l|}{50 neutral + 20 angry + 20 happy}          & \multicolumn{1}{c|}{0.31\%}           & -1.53\%              & \multicolumn{1}{c|}{2.71\%}             & 5.80\%               & 11.34\%              & -2.05\%              & 3.00\%               \\
\bottomrule
\end{tabular}
\vspace{-0.3cm}
\label{tab:synthetic_192}
\end{table*}

%% file: synthetic_utterance_performance_all.tex
\begin{table*}[h]
\centering
\caption{Relative improvement in EER comparing the SV models trained with utterances directly collected from speakers (baseline) and with addition of synthetic angry and happy utterances. Each speaker contains varying numbers of neutral and emotional utterances. Positive values signify an improvement in performance compared to the baseline.}

\vspace{0.2cm}
\small
\setlength{\tabcolsep}{2.5pt}
\begin{tabular}{ll|c|cc|cccc|c}
\toprule
\multicolumn{2}{c|}{\textbf{Experiment Configuration}}                                                                                                        & \textbf{Overall} & \textbf{Neutral} & \textbf{Emotional} & \textbf{Happy} & \textbf{Angry} & \textbf{Sad} & \textbf{Calm} & \textbf{\begin{tabular}[c]{@{}c@{}}Performance gap \\ (Emotional - Neutral)\end{tabular}} \\ \midrule
\multicolumn{1}{l}{\textbf{Baseline}}                                                                                       & No synthetic data & -                & -                & -                 & -              & -           & -            & -             & 1.30\%                                                                                       \\ \midrule
\multicolumn{1}{l}{\multirow{3}{*}{\textbf{\begin{tabular}[c]{@{}l@{}}Synthetic angry + \\ happy utterances\end{tabular}}}} & 5 angry + 5 happy   & 0.45\%           & 0.81\%           & 1.08\%             & 6.16\%         & 1.87\%         & 0.90\%       & 1.54\%        & 1.27\%                                                                                       \\
\multicolumn{1}{l}{}                                                                                                        & 10 angry + 10 happy & -0.31\%          & -0.65\%          & 2.63\%             & 6.16\%         & 1.87\%         & 0.60\%       & -1.02\%       & 1.05\%                                                                                       \\
\multicolumn{1}{l}{}                                                                                                        & 15 angry + 15 happy & 0.00\%           & -1.18\%          & 3.64\%             & 6.16\%         & 1.87\%         & 0.90\%       & -0.51\%       & 0.94\%                                            \\
\bottomrule
\end{tabular}
\vspace{-0.3cm}
\label{tab:synthetic_all}
\end{table*}